\documentclass[prl,twocolumn]{revtex4}
\usepackage{graphicx}
\usepackage{amsmath}
\usepackage[utf8]{inputenc}
\usepackage[english]{babel}

\newcommand{\Tr}{\,\mathrm{Tr}}

\begin{document}

\title{Free energy fluctuations and chaos in the Sherrington-Kirkpatrick model}
\author{T. Aspelmeier}
\affiliation{Max Planck Institute for Dynamics and Self Organization, Göttingen,
Germany}
\pacs{75.50.Lk,75.10.Nr}

\begin{abstract}
The sample-to-sample fluctuations $\Delta F_N$ of the free energy in the
Sherrington-Kirkpatrick model are shown rigorously to be related to
bond chaos. Via this connection, the fluctuations become analytically
accessible by replica methods. The replica calculation for bond chaos shows
that the exponent $\mu$ governing the growth of the fluctuations with system
size $N$, $\Delta F_N\sim N^\mu$, is bounded by $\mu\le\frac 14$.
\end{abstract}

\maketitle

The sample-to-sample fluctuations of the free energy in the mean-field Ising spin
glass \cite{Sherrington:1975} are a long standing unsolved problem in spin glass
physics. In addition to their intrinsic interest as a finite size effect in spin glasses, they are of
fundamental importance for the physics of finite-dimensional spin glasses. It
has been shown \cite{Aspelmeier:2003a} that the finite-size scaling of the free
energy fluctuations $\Delta F_N$ in the mean-field spin glass is equal to the
scaling of the domain wall energy $\Delta F_{DW}$ in finite dimensions $d\ge 6$,
i.e.\ $N^\mu\propto \Delta F_N \propto \Delta F_{DW}\propto L^\theta$, where $N$
is the total system size and $L$ its linear dimension (in the case of a finite
dimensional system). This implies the relationship $\theta=d\mu$ between the
domain wall exponent $\theta$ and the fluctuation exponent $\mu$. This highly
nontrivial equivalence between a mean-field quantity and a finite-dimensional
quantity provides a strong test for replica field theory which was used in
Ref.~\onlinecite{Aspelmeier:2003a} to derive this result.

Chaos is also a very important aspect of spin glasses. Chaos refers to the
property that an infinitesimal change of, for instance, the temperature or the
bond strengths results in a complete change of the equilibrium state. Chaos was
first suggested in the context of the droplet picture and finite dimensional spin
glasses \cite{Bray:1987} but has also been studied in the mean field model
\cite{Rizzo:2001,Rizzo:2003,Krzakala:2005}.

In this paper we derive a new and exact connection between the free energy
fluctuations and the seemingly unrelated phenomenon of chaos. Such a connection
has been suggested by Bouchaud et al.\ \cite{Bouchaud:2003} as part of a
heuristic argument to obtain the free energy fluctuations. Our results partly
corroborate the argument but we will see that a crucial ingredient seems to be
missing from it. In addition to making the heuristic argument precise, our results
provide a new way to access the fluctuations analytically. The fluctuations are a
subextensive quantity such that their calculation usually requires higher order
terms in the loop expansion. These are, however, inaccessible due to the massless
modes present throughout the spin glass phase. Here we will show that it is
sufficient to calculate chaos to zero loop order (which is possible) to obtain
the fluctuations. We demonstrate this explicitly above and at the critical
temperature but believe and present evidence that it also works in the low
temperature phase.

The method we use to derive the connection between fluctuations and chaos is a
variation of the interpolating Hamiltonian method. Our approach is inspired by
the work of Billoire \cite{Billoire:2006} where a similar method was introduced
to study the finite size corrections to the free energy numerically. In this
paper, we will set up a general formalism which will be amenable to an analytical
treatment and derive the upper bound $\mu\le\frac 14$. The details of the
calculation will be published elsewhere \cite{Aspelmeier:2008a}.

Numerically, the sample-to-sample fluctuations in the mean-field model have been
investigated extensively in the literature
\cite{Cabasino:1988,Bouchaud:2003,Palassini:2003,Boettcher:2005a,Katzgraber:2005,Pal:2006}.
In all of these cases the study was restricted to zero temperature. The
fluctuation exponent appears to be $\mu\approx 0.25$, although other values of $\mu$ can not
entirely be ruled out. This value of $\mu$ is also supported by heuristic
arguments put forward in \cite{Aspelmeier:2003a,Bouchaud:2003}. However,
$\mu=\frac 14$ would violate the relation $\theta=d\mu$ since the numerical
results for $\theta$ in high dimensions by Boettcher
\cite{Boettcher:2004a,Boettcher:2004b} give for instance $\theta=1.1\pm 0.1$ for
$d=6$ while $d\mu=1.5$ for $\mu=0.25$. It is not entirely clear whether the
exponent $\mu$ is the same at $T=0$ and at finite temperature. While it most
likely is identical, this is very hard to check since it is difficult to
calculate $\mu$ numerically at any finite temperature. With the connection to
chaos, however, it will be possible in the future to calculate $\mu$ at finite
temperature by simulating chaos. Although temperature chaos is difficult to
simulate as it is a tiny effect, bond chaos, which is relevant here, is much
stronger and is visible more easily \cite{Krzakala:2005}.

Other values than $0.25$ for $\mu$ have been put forward in the literature.
Crisanti et al.\ \cite{Crisanti:1992} found, using a result at zero-loop order by
Kondor \cite{Kondor:1983b}, that $\mu=\frac 16$. The argument is however not
entirely rigorous. Nevertheless, it has recently been argued by a combination of
heuristic arguments and extensive numerical simulations at finite temperature
that $\mu=\frac 16$ is indeed correct \cite{Aspelmeier:2007}. The bound
$\mu\le\frac 14$ derived here is compatible with this but, unfortunately, does
not rule out $\mu=\frac 14$.

Analytically, the free energy fluctuations of any disordered system can in
principle be found with the replica method. Given the partition function $Z$ of a
system of size $N$, it can easily be shown that a Taylor expansion of $\log
\overline{Z^n}$ in powers of $n$ yields $\log\overline{Z^n} = -n\beta F_N +
\frac{n^2}{2}\Delta F_N^2 + \cdots$, where the overbar means the average over the
disorder, $\beta=1/k_B T$ is the inverse temperature and $F_N$ is the average
free energy at system size $N$. The dots indicate higher order cumulants. Using
the replica formalism, one can calculate $\overline{Z^n}$ for integer $n$ and try
to continue the resulting expression to real (or, indeed, complex) $n$ and
isolate the coefficient of the second order term which represents the
fluctuations. In the case of the Ising spin glass this works very nicely above
and at the critical temperature. It is straightforward to show with the standard
replica formalism for the mean-field spin glass \cite{Mezard:1987} that in the
high temperature phase ($\beta<1$), where the saddle point is replica symmetric
and its Hessian has only strictly positive eigenvalues, the fluctuations are
\begin{align}
\beta^2\Delta F_N^2&=-\frac 12 
\log(1-\beta^2)-\frac{\beta^2}{2}+\mathcal O(1/N)
\label{fluctabove}
\end{align} 
\cite{Crisanti:1992,Parisi:1993}. As the critical temperature $T_c$ is
approached ($\beta\nearrow 1/T_c=1$), this expression diverges, which indicates that the 
fluctuations at the critical point must also diverge with $N$. A 
straightforward extension of the calculation in \cite{Parisi:1993} shows that 
the fluctuations at the critical point are 
\begin{align}
\beta^2\Delta F_N^2&=\frac 16 \log N 
+ \mathcal O(1),
\label{fluctat}
\end{align}
which does indeed diverge as $N\to\infty$.
 
Note that Eq.~\eqref{fluctabove} is a one-loop result. Eq.~\eqref{fluctat} even requires
reorganization of the perturbation series \cite{Parisi:1993}. We will see below
that we can obtain precisely the same results from a zero-loop order calculation
of bond chaos.

\paragraph{Interpolating Hamiltonians.}
In order to derive the connection between the fluctuations and chaos, we need to
introduce the following interpolating Hamiltonians:
\begin{align}
  \mathcal H_t^{(r)} &= -\sqrt{\frac{1-t}{N}}\sum_{i<j}J_{ij}s_i s_j - 
  \sqrt{\frac{t}{N}}\sum_{i<j}J^{(r)}_{ij}s_i s_j 
\label{hamiltonians}
\end{align}
with $N$ Ising spins $s_i$, $0\le t\le 1$, $r=1,2$ and $J_{ij}$, $J^{(1)}_{ij}$,
$J^{(2)}_{ij}$ independent Gaussian random variables with unit variance. The
parameter $t$ interpolates between one spin glass system ($t=0$) and a statistically independent, but
otherwise identical one at $t=1$. It is important to note that also for each
other value of $t$ the Hamiltonians describe a normal spin glass, the coupling
constants being $\sqrt{1-t}J_{ij}+\sqrt{t}J^{(r)}_{ij}$  which are Gaussian
random variables of unit variance.

The partition functions of these Hamiltonians are $Z^{(r)}_t=\Tr\exp(-\beta
\mathcal H_t^{(r)})$. Denoting the
average over all coupling constants $J_{ij}$, $J_{ij}^{(1)}$ and $J_{ij}^{(2)}$
by $E_J$, it is straightforward to show that
\begin{align}
E_{J}(\log Z_1^{(1)}-\log Z_0^{(1)})^2 &= 2\beta^2\Delta F_N^2 
 \label{interpolation1}\\
 E_J(\log Z_1^{(1)} - \log Z_0^{(1)})(\log Z_1^{(2)} - \log Z_0^{(2)}) &=
 \beta^2 \Delta F_N^2.
\end{align}
This gives us two distinct representations of the fluctuations. Using the idea
from \cite{Guerra:2002} to represent $\log Z_1^{(r)} - \log Z_0^{(r)}$ by
differentiating with respect to the interpolation parameter and immediately 
integrating again, the
fluctuations can be written in two ways as
\begin{align}
  \beta^2\Delta F_N^2 &= \frac 12 \int_0^1 dt\int_0^1
  d\tau\,E_{J}\frac{\partial\log
  Z_t^{(1)}}{\partial t}\frac{\partial\log Z_\tau^{(1)}}{\partial \tau}
  \label{route1}\\
  &= \int_0^1 dt\int_0^1 d\tau\,E_J \frac{\partial\log
  Z_t^{(1)}}{\partial t}\frac{\partial\log Z_\tau^{(2)}}{\partial \tau}.
  \label{route2}
\end{align}

In \cite{Aspelmeier:2008a} it will be shown how to manipulate this expression in
order to arrive at the following Eqs.~\eqref{average4a} and \eqref{average4b}.
\begin{align}
E_{J}\frac{\partial\log Z_t^{(1)}}{\partial t}\frac{\partial\log
Z_\tau^{(1)}}{\partial \tau} &= \frac{N^2\beta^4}{16} h(t,\tau)
       E_{J}
      \langle(q_{13}^2-q_{14}^2)(q_{13}^2-q_{23}^2)\rangle\nonumber\\
       &\quad + \frac{N \beta^2}{4\sqrt{t\tau}} \left(E_{J} \langle q_{13}^2\rangle - \frac 1N\right)
     \label{average4a} \\
E_J \frac{\partial\log Z_t^{(1)}}{\partial t}\frac{\partial\log
Z_\tau^{(2)}}{\partial \tau} &= \frac{N^2\beta^4}{16}     
      E_{J} \langle(q_{13}^2-q_{14}^2)(q_{13}^2-q_{23}^2)\rangle\nonumber\\
       &\quad + \frac{N \beta^2}{8\sqrt{1-t}\sqrt{1-\tau}} \left(E_{J}
       \langle q_{13}^2\rangle - \frac 1N\right).
     \label{average4b}
\end{align}
with $h(t,\tau)=2-\frac{\sqrt{1-t}\sqrt{\tau}}{\sqrt{t}\sqrt{1-\tau}}
-\frac{\sqrt{1-\tau}\sqrt{t}}{\sqrt{\tau}\sqrt{1-t}}$.
Note that these equations are exact.
The symbols $q_{ab}$ are overlaps between independent replicas with different
interpolation parameters,
\begin{align}
q_{ab}(t,\tau) &= \frac 1N \sum_i s_i^{a,t} s_i^{b,\tau}.
\end{align}
Although Eqs.~\eqref{average4a} and \eqref{average4b} are formally very similar,
there is an important difference. In Eq.~\eqref{average4a} replicas $1$ and $2$
have Hamiltonian $\mathcal H_t^{(1)}$ and replicas $3$ and $4$ have Hamiltonian
$\mathcal H_\tau^{(1)}$. In Eq.~\eqref{average4b}, on the other hand, replicas
$1$ and $2$ have Hamiltonian $\mathcal H_t^{(1)}$ while replicas $3$ and $4$
have Hamiltonian $\mathcal H_\tau^{(2)}$. The angular brackets
$\langle\cdots\rangle$ denote the thermal average of a system of independent
replicas with the appropriate Hamiltonians.

The last important step is to employ the fact that for any given value of $t$,
$\mathcal H_t^{(r)}$ represents a normal mean-field spin glass with
Gaussian couplings just like any other. Consider Eq.~\eqref{average4a}. The overlap
$q_{13}$ between two replicas with different interpolation parameters is nothing
but the overlap between two normal spin glasses with identical bonds (if
$t=\tau$), uncorrelated bonds (if $t=0, \tau=1$ or vice versa) or related, but
not equal bonds (for anything in between). Similarly, for Eq.~\eqref{average4b}
the overlap is between systems with equal bonds ($t=\tau=0$), totally
uncorrelated bonds ($t=1$ or $\tau=1$) or correlated bonds (anything else).
This shows the connection to bond chaos.

The overlaps in Eqs.~\eqref{average4a} and \eqref{average4b} thus do not depend
on $t$ and $\tau$ separately but only on a measure of ``distance'' $\epsilon$ between the two sets
of bonds. We define $\epsilon$ via the correlation between the bonds, i.e.\ we
set $\frac{1}{\sqrt{1+\epsilon^2}}=E_J
(\sqrt{1-t}J_{ij}+\sqrt{t}J_{ij}^{(1)})(\sqrt{1-\tau}J_{ij}+\sqrt{\tau}J_{ij}^{(1)})
= \sqrt{1-t}\sqrt{1-\tau}+\sqrt{t\tau}$ for Eq.~\eqref{average4a} and
$\frac{1}{\sqrt{1+\epsilon^2}}=E_J(\sqrt{1-t}J_{ij}+\sqrt{t}J_{ij}^{(1)})(\sqrt{1-\tau}J_{ij}
+\sqrt{\tau}J_{ij}^{(2)})=\sqrt{1-t}\sqrt{1-\tau}$ for Eq.~\eqref{average4b}.
With these definitions, $\epsilon=0$ means identical bonds and $\epsilon=\infty$ means totally
uncorrelated bonds.

We can now make a change of variables under the integrals in
Eqs.~\eqref{route1} and \eqref{route2} and eliminate, say, $\tau$ in favor of
$\epsilon$. The remaining integral over $t$ can be carried out analytically and
we get the two different exact expressions
\begin{align}
\beta^2\Delta F_N^2 &= -\frac{N^2\beta^4}{16}\int_0^\infty d\epsilon \,
f_1(\epsilon) E_J
\langle(q_{13}^2-q_{14}^2)(q_{13}^2-q_{23}^2)\rangle\nonumber\\ &\quad +
\frac{N \beta^2}{4}\int_0^\infty d\epsilon\,g_1(\epsilon)\left(E_J
     \langle q_{13}^2\rangle - \frac 1N\right)
     \label{fluct1} \\
     &= +\frac{N^2\beta^4}{16}\int_0^\infty d\epsilon \,
f_2(\epsilon) E_J
\langle(q_{13}^2-q_{14}^2)(q_{13}^2-q_{23}^2)\rangle\nonumber\\ &\quad +
\frac{N \beta^2}{4}\int_0^\infty d\epsilon\,g_2(\epsilon)\left(E_J
     \langle q_{13}^2\rangle - \frac 1N\right)
     \label{fluct2} 
\end{align}
where
\begin{align}
f_1(\epsilon) &= 
\frac{4\epsilon^2\arcsin\frac{1}{\sqrt{1+\epsilon^2}}}{(1+\epsilon^2)^2},
&
g_1(\epsilon) &=
\frac{2\arcsin\frac{1}{\sqrt{1+\epsilon^2}}}{(1+\epsilon^2)^{3/2}}, \\
f_2(\epsilon) &= \frac{2\epsilon\log(1+\epsilon^2)}{(1+\epsilon^2)}, &
g_2(\epsilon) &= \frac{\epsilon\log(1+\epsilon^2)}{(1+\epsilon^2)^{3/2}}.
\end{align}
By going over from $t$ and $\tau$ to $\epsilon$ the distinction between the
different choice of Hamiltonians in the two representations of the fluctuations
has disappeared and the overlaps as a function of $\epsilon$ in
both of these equations are the same.

Note the minus sign in front of the first term in Eq.~\eqref{fluct1} as opposed
to the plus sign in Eq.~\eqref{fluct2}. Since the function $f_1(\epsilon)$ is
nonnegative, the first term is indeed a negative contribution. We conclude that
the second term in Eq.~\eqref{fluct1} is an upper bound for the fluctuations.

\paragraph{Probability distribution of the overlap.}
If we had the disorder averaged probability distribution $P_\epsilon(q)$ to find
the overlap $q$ between two replicas with bond distance $\epsilon$, we could
evaluate $E\langle q_{13}^2(\epsilon)\rangle$. In order to evaluate
$E\langle(q_{13}^2-q_{14}^2)(q_{13}^2-q_{23}^2)\rangle$ we need the probability
distribution $P_\epsilon^{123}(q_{13},q_{23})$ to simultaneously find $q_{13}$
and $q_{23}$, as well as the probability distribution
$P_\epsilon^{1234}(q_{14},q_{23})$ to find $q_{14}$ and $q_{23}$. However, in
this paper we will focus on $P_\epsilon(q)$.

The probability distribution $P_\epsilon(q)$ can be calculated approximately
from large deviation statistics principles by considering two replicas with
bonds $J_{ij}^0$ and $J_{ij}(\epsilon)$ which are a bond distance $\epsilon$ apart
and constraining their overlap to a given value of $q$. The partition
function $Z_{\epsilon,J}(q)$ of this combined system is
\begin{align}
Z_{\epsilon,J}(q) &= \mathrm{Tr}\,\delta\left(q-\frac{\sum_i s_i
t_i}{N}\right)\exp\left( \beta \sum_{i<j}\left(J_{ij}^0 s_i s_j + 
J_{ij}(\epsilon) t_i t_j\right) \right).
\label{constrained}
\end{align}
The variables $s_i$ and $t_i$ are the spin variables of the two replicas. 
From this one gets the average free
energy per spin $\beta f_{\epsilon}(q) = -\frac 1N E_J \log Z_{\epsilon,J}(q)$
and $P_\epsilon(q)$ is approximated by
\begin{align}
P_\epsilon(q) &\approx P_\epsilon^0(q) := \frac{e^{-N\beta
f_\epsilon(q)}}{\int_0^1 dq\,e^{-N\beta f_\epsilon(q)}}.
\label{P0}
\end{align}
Averages over $P_\epsilon^0(q)$ will be denoted by $[\cdots]_0$.

A more precise discussion of the finite size effects and the relevance for
``small'' deviations will be given in \cite{Aspelmeier:2008a}. Here we will
show that $P_\epsilon^0(q)$ is indeed the correct probability distribution to
use by demonstrating that it yields the exactly known results above and at the critical
temperature and by comparing predictions in the spin glass phase with
simulations from \cite{Krzakala:2005} (see below).

\paragraph{Replica calculation.}
Temperature chaos in mean-field spin glasses has been treated in the literature
\cite{Rizzo:2001,Rizzo:2003}. However, to the best of our knowledge, bond chaos
has never been calculated and we will therefore present a brief sketch of
our results here. Since these replica calculations are fairly standard, we refer
the reader to \cite{Rizzo:2001,Rizzo:2003} for details. Repeating Rizzo's
calculation \cite{Rizzo:2001} for the constrained two-replica partition
function from Eq.~\eqref{constrained} but for bond chaos rather than temperature
chaos, one arrives at the following truncated replica free energy
\begin{widetext}
\begin{align}
\beta f_\epsilon(q) &= qp_d-\frac{p_d^2}{2} - \frac{yp_d^4}{6} -
\frac{q^2}{2(1-2\tau')} + \tau\int_0^1 dz\,q^2(z) + \tau'\int_0^1 dz\,p^2(z) +
\frac y6 \int_0^1 dz\,\left(q^4(z) + p^4(z)\right)
 -\frac w3\int_0^1 dz\,zq^3(z)\nonumber\\
&\quad-w\int_0^1 dz\,zp^2(z)q(z) - w\int_0^1
dz\int_z^1 dz'\,\left((q^2(z)+p^2(z))q(z') + 2p(z)q(z)p(z')\right)
+2wp_d\int_0^1 dz\,p(z)q(z).
\label{free}
\end{align}
\end{widetext}
The parameters $w$ and $y$ are equal to $1$ for the SK model and
$\tau=(\beta^2-1)/(2\beta^2)$ is the distance from the critical temperature
$T_c=1/\beta_c=1$. The only reference to the bond distance is contained in
$\tau'=(\beta^2-\sqrt{1+\epsilon^2})/(2\beta^2)$. Three saddle point
equations can be derived from this free energy by differentiating with respect to
$q(x)$, $p(x)$ and $p_d$. The function $q(x)$ is the Parisi function for the
overlap of the first replica with itself (the same function of course applies by
symmetry to the overlap of the second replica with itself). The function $p(x)$
describes the overlap between replica one and two. The parameter $p_d$ stems from
the diagonal of the overlap matrix between replicas one and two and is a
conjugate variable to the forced overlap $q$.

Solving the saddle point equations is nontrivial and only possible in certain
limiting cases. Deferring the details to \cite{Aspelmeier:2008a}, we summarize
the results here. Above the critical temperature we find 
$\beta f_\epsilon(q) = \frac{q^2}{2}\left(
1-\frac{\beta^2}{\sqrt{1+\epsilon^2}}\right) + \mathcal O(q^4)$. 
With this we can calculate $P^0_\epsilon(q)$ according to Eq.~\eqref{P0}.
Above the critical temperature, there is no replica symmetry breaking and the
distributions $P_\epsilon^{123}(q_{13},q_{23})$ and
$P_\epsilon^{1234}(q_{13},q_{24})$ factorize into a product of
$P_\epsilon^0(q)$'s. The expression $E_J\langle
(q_{13}^2-q_{14}^2)(q_{13}^2-q_{23}^2)\rangle$ can thus be written as
$[q^4]_0-[q^2]_0^2$. The integrals in Eq.~\eqref{fluct2} can then be
evaluated exactly giving precisely the result from Eq.~\eqref{fluctabove}.

Exactly at the critical temperature there is still no replica symmetry breaking
and the factorization of $P_\epsilon^{123}$ and $P_\epsilon^{1234}$ applies as
above. We obtain $P_\epsilon^0(q)\sim e^{-N w q^3/6}$ for $\epsilon\ll
N^{-1/6}$ and $P_\epsilon^0(q)\sim e^{-N q^2\epsilon^2/4}$ for
$N^{-1/6}\ll \epsilon\ll 1$. These limiting cases are enough to calculate the
leading behavior of the integrals in Eq.~\eqref{fluct2} and we get the same as
in Eq.~\eqref{fluctat}.

Below the critical temperature, replica symmetry breaking does apply and we can
not factorize $P_\epsilon^{123}$ and $P_\epsilon^{1234}$. Although it has been
shown in \cite{Parisi:2000,Guerra:1996} how to break down these probability
distributions, the results only apply for $\epsilon=0$. We therefore concentrate
on the second integral in Eq.~\eqref{fluct1} which is an upper bound for the
fluctuations. We find four regimes,
\begin{align}
P_\epsilon^0(q)&\sim \left\{
\begin{array}{l@{\hspace{1cm}}l}
1 & \epsilon\ll N^{-1/2} \\
e^{-Nc_1q^3\epsilon^2} & N^{-1/2}\ll \epsilon\ll N^{-1/5} \\
e^{-Nc_2q^2\epsilon^3} & N^{-1/5}\ll \epsilon\ll 1 \\
e^{-Nq^2f(\epsilon)} & \text{otherwise}
\end{array}
\right. ,
\label{Pbelow}
\end{align}
where $c_{1,2}$ are constants and $f(\epsilon)$ is an (unimportant) function.
Using these results we can calculate the leading behavior of $\int_0^\infty d\epsilon E_J\langle
q_{13}^2\rangle$ and obtain for the dominant contribution to the integral
\begin{align}
\frac{N \beta^2}{4}\int_0^{N^{-1/5}}
d\epsilon\,g_1(\epsilon) [q^2]_0 &=
N^{1/2}\int_0^{N^{3/5}} dx\,\mathcal F(x)
\sim N^{1/2},
\label{fluctbelow}
\end{align}
where $x=\epsilon\sqrt{N}$ is a scaling variable and $\mathcal F(x)$ is a
scaling function with the properties $\mathcal F(x)\to\text{const.}$ ($x\to 0$) and $\mathcal F(x)\sim x^{-4/3}$
($x\to\infty$).  This behavior of $\mathcal F(x)$ is perfectly
consistent with the numerical results for $[q]_0$ presented in
\cite{Krzakala:2005}. This is strong evidence that the finite size corrections
are indeed irrelevant even in the low temperature phase. (Note
$x'=N\epsilon^2$ is used as a scaling variable in \cite{Krzakala:2005} and
the scaling of $[q]_0$ is investigated instead of $[q^2]_0$ as we do here. This
is however only a trivial difference. Although they indicate in their scaling
plot Fig.~1 a decrease proportional to ${x'}^{-1/2}$, visual inspection of that 
plot shows that a slower decrease $\sim {x'}^{-1/3}$ is more likely which
would coincide with the scaling of $x^{-4/3}$ for $\mathcal F(x)$ here.)

Since Eq.~\eqref{fluctbelow} is an upper
bound for the fluctuations, it follows that $\mu\le \frac 14$.

\paragraph{Discussion.}
We have shown that there exists a deep and exact relation between the free energy
fluctuations and chaos in spin glasses, Eqs.~\eqref{fluct1} and \eqref{fluct2}. A
similar connection has been suggested by Bouchaud et al.\ in a heuristic argument
\cite{Bouchaud:2003}. Briefly, the argument runs as follows. When the bonds are
changed randomly by an amount of order $1/\sqrt{N}$, the ground state of the
system changes and we get a new ground state energy. In order to obtain a truly
independent new bond configuration, we must change the bonds by an amount of
order one., i.e.\ we obtain a sequence of $\sim\sqrt{N}$ level crossings. Each
of these contributes a random amount of order $1$ to the change in ground state
energy such that the final energy differs by an amount of order $N^{1/4}$ from
the original ground state energy, i.e.\ $\mu=\frac 14$.

This argument corresponds precisely to the second term in Eq.~\eqref{fluct1}.
From Eq.~\eqref{Pbelow} we see that the overlap does not change appreciably until
$\epsilon\approx N^{-1/2}$ (corresponding to the statement that the ground state
does not change up to that value of $\epsilon$), and the leading contribution to
the fluctuations comes from tuning $\epsilon$ from $0$ to $\mathcal O(1)$ and is
proportional to $N^{1/4}$. There is, however, the negative first term in
Eq.~\eqref{fluct1} which has no correspondence in the heuristic argument and
which reduces the size of the fluctuations and might decrease $\mu$. We are
currently unable to provide an intuitive explanation for this term. We note,
however, that it is crucial and can not be ignored since in our second
formulation, Eq.~\eqref{fluct2}, the integral containing the four replica
overlaps is the dominant term and is solely responsible for $\mu$. A deeper
understanding of these matters would be highly desirable.

\paragraph{Acknowledgments.}
I would like to thank M.\ Goethe and M.A.\ Moore for many interesting and useful
discussions.

\bibliography{LiteraturDB,cond-mat}

\end{document}